\documentclass[12pt, preprint]{aastex}
\begin{document}

\title{Interferometric $890\micron$ Images of High Redshift Submillimeter Galaxies}
\author {D. Iono\altaffilmark{1,2}, A. B. Peck\altaffilmark{1}, 
A. Pope\altaffilmark{3}, C. Borys\altaffilmark{4}, 
D. Scott\altaffilmark{3}, D. J. Wilner\altaffilmark{1}, 
M. Gurwell\altaffilmark{1}, P. T. P. Ho\altaffilmark{1},
M. S. Yun\altaffilmark{5}, 
S. Matsushita\altaffilmark{6},
G. R. Petitpas\altaffilmark{1}, 
J. S. Dunlop\altaffilmark{7}, 
M. Elvis\altaffilmark{1}, 
A. Blain\altaffilmark{4},
E. Le Floc'h\altaffilmark{8}}
\altaffiltext{1}{Harvard-Smithsonian CfA, 60 Garden St., Cambridge, MA 02138}
\altaffiltext{2}{National Astronomical Observatory of Japan, 2-21-1 Osawa, Mitaka, Tokyo 181-8588; d.iono@nao.ac.jp}
\altaffiltext{3}{Dept. of Physics \& Astro., UBC, Vancouver, BC, V6T 1Z1}
\altaffiltext{4}{California Institute of Technology, Pasadena, CA 91125}
\altaffiltext{5}{Dept. of Astronomy, Univ. of Mass., Amherst, MA 01003}
\altaffiltext{6}{Academia Sinica Institute of Astronomy and Astrophysics, P.O. Box 23-141, Taipei 106, Taiwan, R.O.C.}
\altaffiltext{7}{Institute for Astronomy, University of Edinburgh, Royal Observatory, Blackford Hill, Edinburgh EH9 3HJ}
\altaffiltext{8}{Steward Observatory, 933 N. Cherry Ave, Tucson, AZ 85721}
\begin{abstract}

We present high resolution $890\micron$ images of two 20~mJy 
submillimeter galaxies, SMMJ123711+622212 
and MIPS~J142824.0+352619, obtained using the   
Submillimeter Array (SMA). 
Using submillimeter interferometric observations  
with an angular resolution of $2\farcs5$, 
the coordinates of these high 
redshift sources are determined with an accuracy of $0\farcs2$. 
The new SMA data on SMMJ123711+622212 
reveal an unresolved submm source
offset to the east by $0\farcs8$ from an optical galaxy 
found in deep $HST$ images, suggesting
either a large galaxy with a dusty central region, or an interacting galaxy system.  
The SMA image of hyper-luminous 
(L$_{\rm FIR} = 3.2 \times 10^{13}$~L$_{\odot}$) 
source MIPS~J142824.0+352619 provides 
a firm upper limit to the source size of 
$\lesssim 1\farcs2$. This constraint 
provides evidence that the foreground lens is only weakly 
affecting the observed high FIR luminosity.

\end{abstract}


\section{Introduction}

The discovery of high redshift submillimeter sources has significantly
improved our understanding of the star formation history in the early universe.
Negative k-correction allows observation of the thermal dust emission
at $850\micron$
almost independent of the redshift up to $z\sim 10$ \citep{blain02}. 
Deep observations using the SCUBA bolometer on the 
James Clerk Maxwell Telescope (JCMT) have 
unveiled the presence of distant submm sources 
\citep{smail97,barger98,hughes98,eales00,cowie02, scott02,borys03, webb03, wang04}. 
The primary origin of the submm emission is believed to be  
the reprocessed dust emission from newborn stars in young galaxies.
While these discoveries are attended by a great number of high resolution 
follow-up optical/NIR imaging studies, the $14''$ 
resolution of the
JCMT at 850$\micron$ yields a large error circle which 
is too coarse for a precise determination of the 
optical/NIR counterparts to these sources.  Deep optical imaging 
typically shows several optical/NIR 
sources within the SCUBA beam.
To date, the most successful ways 
to obtain precise astrometry on the target are to obtain high 
resolution, deep 1.4~GHz radio images  
\citep{ivison98,barger00,chapman01,ivison02,dunlop04},
or to obtain interferometric 1.3~mm continuum images \citep[e.g.][]{downes99}. 
The former, however, does not identify
robust optical/NIR counterparts for all of the sources, revealing counterparts for 
$\sim75\%$ of the $S_{850\micron} > 5$~mJy sources with  
$S_{\rm 1.4GHz} > 30$~$\mu$Jy 
\citep[e.g.][]{ivison02,chapman03,borys04,greve04,wang04}.  
Precise astrometry obtained using mm or submm interferometers 
can unambiguously identify the correct counterpart
for the remaining radio-faint sources,
and sources with multiple radio counterparts.
High angular resolution submm observations also allow us
to understand the true nature of the submm 
sources with established optical counterparts 
in which gravitational lensing is a possibility.
  
We present recent Submillimeter Array
\citep[SMA;][]{ho04} detections of two $S_{850\micron} \sim 20$~mJy  
sources, SMMJ123711+622212 \citep[hereafter GN~20;][]{pope05} 
and MIPS~J142824.0+352619 
\citep[hereafter MIPS-J1428;][]{borys05}.
The 20.3~mJy source GN~20 was discovered in the recent SCUBA
observations of the GOODS North Field \citep{giavalisco04}.
GN~20, with its $10\sigma$ 
$850\micron$ detection \citep{pope05} and 
$5\sigma$ 1.3~mm detection at IRAM PdBI 
(Pope et al. in preparation) make this one of the 
strongest submm sources discovered to date. 
It has very weak radio and undetectable $450\micron$ emission, 
suggesting it lies at high redshift. 
MIPS-J1428 was discovered in the \textit{Spitzer} MIPS images of the
NDWFS Bo\"{o}tes field \citep{jannuzi99,soifer04}.
It was detected  \citep{borys05} at  $350\micron$ 
using SHARC-II \citep{dowell03} on the CSO, 1.4~GHz radio 
continuum at the VLA, and 
subsequent followup observations \citep{borys05} revealed that it is an 
extremely luminous ${\rm (3.2\pm0.7)\times10^{13}\,L_\odot}$
starburst dominated galaxy at $z=1.325$.
Recent Keck-DEIMOS spectroscopy has revealed a $z=1.034$ galaxy 
directly aligned with MIPS-J1428 \citep{borys05}, possibly lensing it.  
High angular resolution submm observations allow us to determine 
whether or not the submm emission is coincident with the 
optical emission, and thereby search for any evidence for amplification
of the FIR luminosity by gravitational lensing.
We adopt $H_0 = 70$~km~s$^{-1}$~Mpc$^{-1}$,
 $\Omega_m$ = 0.3, $\Omega_{\Lambda}$ = 0.7. 

\section{Observation and Data Reduction}

GN~20 was observed on February 20 (track~1) and 
March 5, 2005 (track~2), and MIPS-J1428  was observed on March 8 (track~3)
 and April 4 (track~4), 2005 using 5 -- 7 antennas in the compact 
configuration of the SMA.  
The SIS receivers were tuned to a center frequency of 342.883~GHz in the
upper sideband (USB),  yielding 332.639~GHz in the lower sideband (LSB). 
This tuning frequency was chosen to facilitate the 
receiver tuning and to achieve the optimal receiver performance. 
The target coordinates were obtained from the  
1.3~mm IRAM PdBI detection of GN~20 ($\alpha$~(J2000)~$= 12^h 37^m 11.88^s$, 
$\delta$~(J2000)~$=62^{\circ}22' 12\farcs00$; Pope et al. in preparation), 
and the position of 
the proposed MIR galaxy of MIPS-J1428 
\citep[$\alpha$~(J2000)~$= 14^h 28^m 24.10^s$, 
$\delta$~(J2000)~$=35^{\circ}26' 19\farcs00$;][]{borys05}.
All tracks were taken under good atmospheric opacity 
(i.e. $\tau_{225} = 0.04$ -- $0.08$).

The SMA data were calibrated using the Caltech software package MIR, 
modified for the SMA.
Antenna based passband calibration was done using 
all of the planets and bright QSOs observed in a given track.
For GN~20, antenna based time-dependent 
phase calibration was done using 1153+495, a 0.7~Jy QSO 14$^\circ$
away from the target.  In addition, a 0.6~Jy QSO near GN~20, 1048+717
(14$^\circ$ from GN~20, 23$^\circ$ from 1153+495) was observed
for a total of 12 minutes during each track.  The detection of 
1048+717 at the phase center empirically verifies and constrains
the accuracy of the 
phase calibration referenced to 1153+495 (see \S3). 
Similarly for the MIPS-J1428 tracks, two QSOs 1310+323 
(0.6~Jy; 16$^\circ$ away from MIPS-J1428) 
and 1635+381 (1.0~Jy; 25$^\circ$ away from MIPS-J1428) were used together 
to calibrate the time dependent phase, 
and 1419+543 (0.4~Jy; 3$^\circ$ away from MIPS-J1428) 
was used to check the astrometry. 
Finally, absolute flux calibration was performed using Callisto and Mars.
Imaging was carried out in MIRIAD \citep{miriad}.   
Maximum sensitivity was achieved by 
adopting natural weighting, which gave a  
synthesized beam size of $2\farcs9  \times 2\farcs2$ (P.A.=
$12.2^\circ$) for GN~20 
and $2\farcs6  \times 2\farcs4$ (P.A. = $-47.2^\circ$) 
for MIPS-J1428.  
The rms noise after combining the two sidebands in two tracks 
was 2.1~mJy (GN~20) and 2.2~mJy (MIPS-J1428).

\section{Results and Discussion}

\subsection{SMMJ123711+622212 (GN~20)}
\begin{figure}
  \plotone{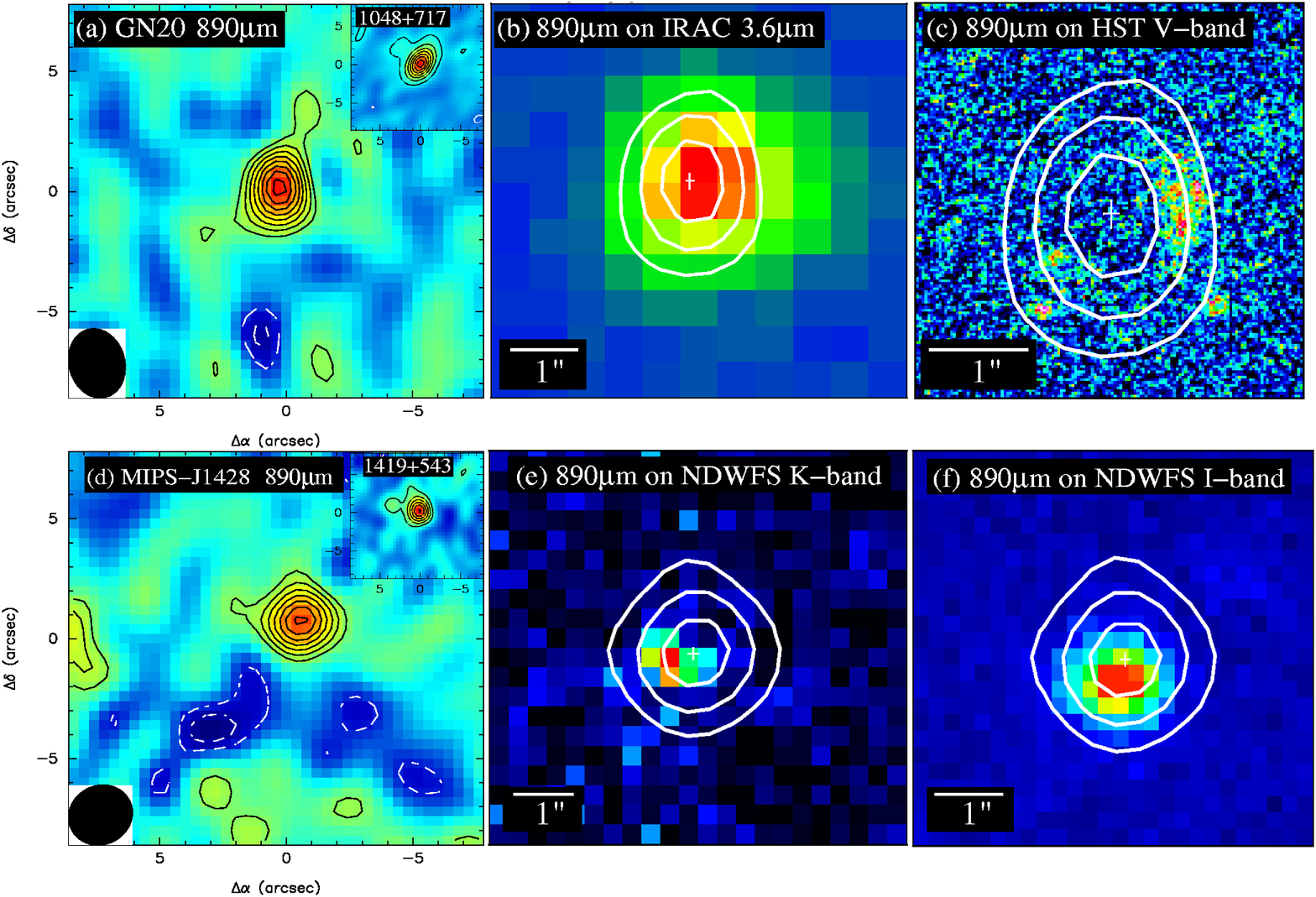}
  \caption{(\textit{a}) The SMA 890$\micron$ map of GN~20 and 
(\textit{inset}) 1048+717.
The lowest positive contours represent $2\sigma$, and the contours increase
by $1 \sigma$ for GN~20, and by $2 \sigma$ for 1048+717.  
The lowest negative contour is 2$\sigma$ and increases by $1 \sigma$.
(\textit{b}) The SMA $890\micron$ contours overlaid on the $Spitzer$
IRAC $3.6\micron$ image and over the (\textit{c}) $HST$ ACS $V$-band image, 
 both obtained from the GOODS archive. 
The IRAC and $HST$ images are corrected for the known $0\farcs38$ offset 
in declination. 
The 4, 6, and 8 $\sigma$ contours from (\textit{a}) are shown, and 
errorbars near the center show the astrometric accuracy of the SMA image.  
(\textit{d}) The SMA map of MIPS-J1428 and (\textit{inset}) 1419+543. 
The contours are the same as in GN~20.
The smallness of the astrometric errors ($0\farcs1$ -- $0\farcs2$) 
from the phase center in the 1048+717 and 1419+543 maps
prove the robustness of the astrometry of GN~20 and MIPS-J1428.
The SMA $890\micron$ map of MIPS-J1428 is shown overlaid on the NDWFS 
(\textit{e}) K-band image and the (\textit{f}) I-band image.
The 3, 5, and 7 $\sigma$ contours from (\textit{d}) are shown, and 
errorbars near the center show the astrometric accuracy of the SMA image.  
}
  \label{fig1}
\end{figure}

\begin{figure}
  \plotone{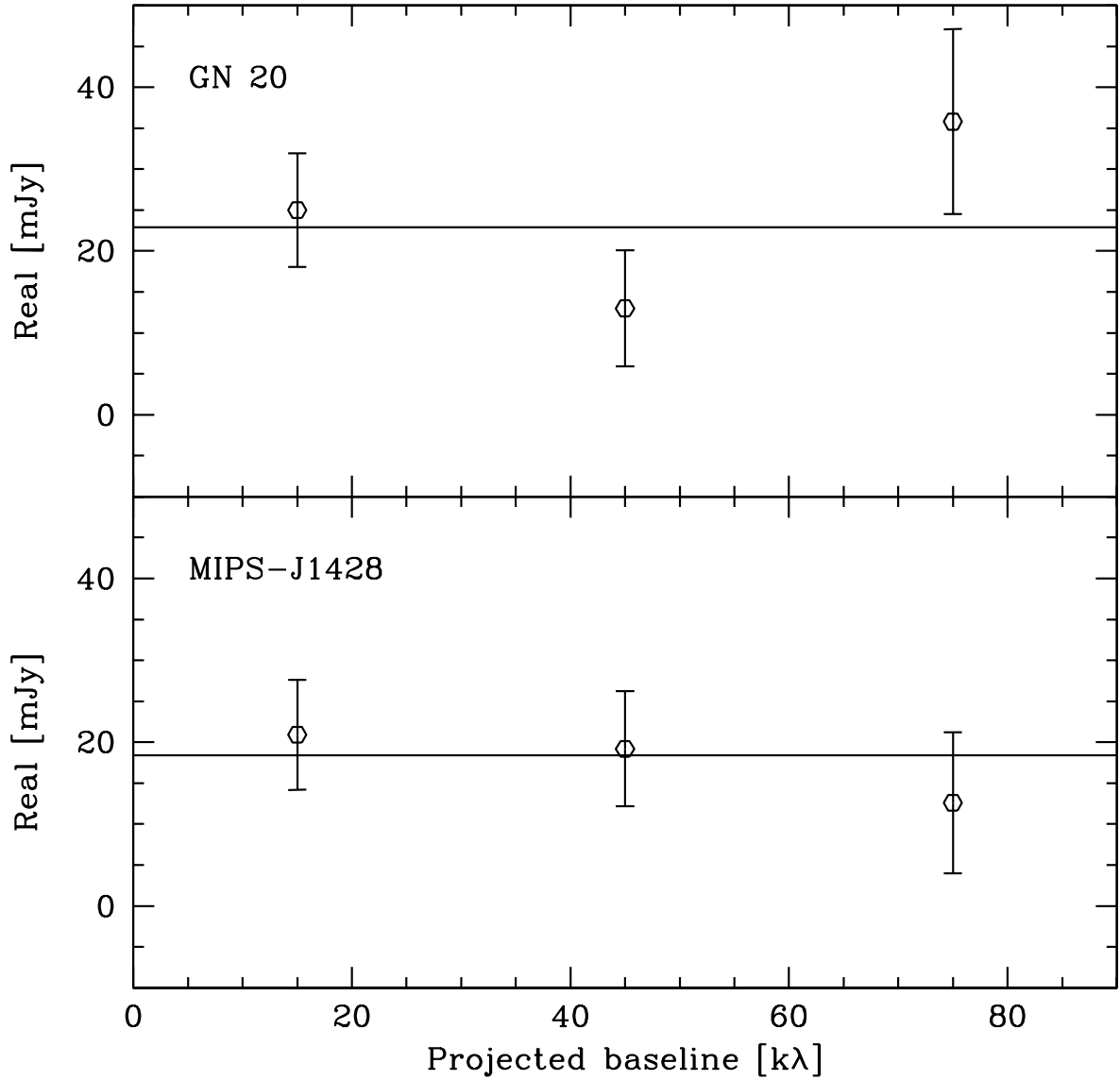}
  \caption{The real part of the visibility amplitudes vs. projected 
baseline length for (\textit{top}) GN~20 and (\textit{bottom}) 
MIPS-J1428.  The solid horizontal lines represent a point source
model with continuum fluxes of 22.9~mJy~(GN~20) and 
18.4~mJy~(MIPS-J1428).  The upper limit to the source sizes of
both of these sources are comparable to the scale constrained 
by the longest baseline length
which is $\sim 1\farcs2$.
}
  \label{fig2}
\end{figure}

The SMA image of GN~20 is shown in Figure~\ref{fig1}~(a).  
The real part of the visibility amplitudes do not decline as a function of
projected baseline length,  
indicating that the emission is not spatially resolved (Figure~\ref{fig2}).
The upper limit on the source size is $1\farcs2$.
The derived total flux from a point source model 
is $22.9 \pm 2.8$~mJy, consistent 
with the $890\micron$ flux of $18.1$~mJy 
extrapolated from the SCUBA $850\micron$ flux of 
$20.3 \pm 2.1$ mJy \citep{pope05}.
The derived coordinates for GN~20 are
$\alpha$~(J2000)~$= 12^h 37^m 11.92^s$, 
$\delta$~(J2000)~$=62^{\circ}22' 12\farcs10$, with 
statistical uncertainties in the fit of  
$0\farcs1$ for both $\alpha$ and $\delta$.  These are consistent with 
estimated errors of $\sim 0\farcs11$ in $\alpha$ and 
$\sim 0\farcs15$ in $\delta$ 
from a $10 \sigma$ detection and a beam of $2\farcs9 \times 2\farcs2$.

In order to check the robustness of our phase calibration and to estimate
the systematic uncertainties in the SMA astrometry,  
we have imaged the test QSO (1048+717)
using the same phase calibration we used to map GN~20.
The resultant QSO map after adding the two sidebands and two tracks
is shown in Figure~\ref{fig1} (a)(inset).
A point source fit to the visibilities of 1048+717 gave a 
positional offset from the phase center of 
$\Delta \alpha = 0\farcs01 \pm 0\farcs02$ and  $\Delta \delta = 0\farcs06 \pm 0\farcs02$. 
The coordinates of this and all of the QSOs used in these observations 
were adopted from the Radio Reference Frame \citep{johnston95},
which are accurate to better than 3 mas.
Hence the precise detection of 1048+717 at the phase center
ensures that our phase calibration referenced to 1154+379 is robust.

As an additional check, we fit a point source model 
to the visibilities of 1048+717 in each sideband of each track separately.
The results show that the offsets from the phase center are
consistent among the different sidebands and tracks in R.A. 
($\Delta \alpha = 0\farcs01$ -- $0\farcs04$),
and slightly larger in Dec. for track~2 ($\Delta \delta = 0\farcs02$ -- $0\farcs05$),
and a factor 10 larger in Dec. for track~1 ($\Delta \delta \sim 0\farcs30$).
The overall average offset from 
the phase center is $\Delta \alpha = 0\farcs02 \pm 0.02$ and 
$\Delta \delta = 0\farcs16 \pm 0.14$.
The uncertainties in Dec. in track~2 are larger than the
uncertainties in R.A. due to 
beam elongation in the north-south direction, while the 
source of the factor 10 larger error in track~1
is not obvious from the data.  
These errors are consistent with the uncertainties of 
$0\farcs1$ -- $0\farcs15$ expected from
a maximum baseline error of $0.1\lambda$ 
at 230~GHz using a phase calibrator 
that is $15^\circ$ away from the target.
These systematic tests prove the robustness of our 
phase calibration in each track
and lend high confidence to the resulting positional accuracy 
of $\lesssim 0\farcs1$ in R.A. and $0\farcs1$ -- $0\farcs2$ in Dec. for GN~20.

The new astrometric coordinates allow us to compare the submm
 source with high resolution images in the publicly available 
deep $Spitzer$ IRAC \citep{fazio04} (Figure~\ref{fig1}~(b)) and 
$HST$ ACS \citep{ford98} (Figure~\ref{fig1}~(c)) images 
of the GOODS North Field.
The absolute astrometric accuracy of both of these images is $\sim 0\farcs1$,
and they are both tied to the coordinate frame defined by the VLA  positions in 
\citet{richards00}\footnote{After applying the known positional offset in Dec.
of $0\farcs38$ between the GOODS-N images and the VLA catalog positions. See 
http://data.spitzer.caltech.edu/popular/goods/Documents/ for details.}.
The IRAC 3.6$\micron$ image reveals a source centered $< 0\farcs5$ west of 
the submm coordinates, while the higher resolution 
$HST$ $V$-band image reveals a faint optical 
source $0\farcs8$ to the west. 
From the analysis of available ACS images, 
it is found that this optical source is a B-dropout galaxy ($B = 27.2 \pm 0.4$,
$V = 25.2 \pm 0.1$, $i = 24.4 \pm 0.1$) which
gives constraints on the probable redshift to be $z\sim 3$--4.
We believe that the $0\farcs8$ offset between the SMA position and the 
\textit{HST} position is significant, and there are several possible
astrophysical explanations for this difference. 
The submm emission may arise 
from part of a large galaxy where $V$-band emission is completely obscured.  
Alternatively, GN~20 might be an interacting
system where the optical galaxy is a companion to the dusty, more actively
star forming galaxy.  Although the observed FIR luminosities are 
significantly different, the apparent separation of 
these galaxies ($\sim 6$~kpc at $z \sim 3 $)
suggest a close similarity to the Antennae system (NGC~4038/39), where
the optical galaxies are separated by 7.5~kpc and 
most of the starburst activity is occurring in the medium between the
two galaxies \citep[e.g.][]{wang04b}.

GN~20 was suggested to be a two component source, GN~20.1 (20.3~mJy) 
and GN~20.2 (11.7~mJy), separated by $18''$~(140~kpc) 
in the low resolution SCUBA image \citep{pope05}.  
The $890\micron$ flux agreement between our SMA observation and 
the SCUBA measurement of GN~20.1 implies that GN~20 may be 
a two component source.  However, the suggested position of GN~20.2 is
beyond the half power point of the SMA primary beam,
where the sensitivity is reduced by more than a factor of two.
Detailed discussion in the context of multi-wavelength
observations of this source 
will be provided in a forthcoming paper (Pope et al. in preparation).

\subsection{MIPS~J142824.0+352619 (MIPS-J1428)}

The SMA map of MIPS-J1428 is shown in Figure~\ref{fig1}~(d).  
The derived total flux is $18.4 \pm 2.5$ mJy and shows 
excellent agreement with the $890\micron$ flux of $19.5$~mJy
extrapolated from the SCUBA $850\micron$ flux of 
$21.9 \pm 1.3$~mJy.
MIPS-J1428 is not spatially resolved with the $2\farcs5$ beam 
(see Figure~\ref{fig2}), 
and the derived coordinates from a point source fit are
$\alpha~(2000)= 14^h 28^m 24.06^s$, $\delta~(2000)=35^{\circ} 26' 19\farcs79$,
with uncertainties in the fit of $0\farcs1$ for both $\alpha$ and $\delta$.
The strong $890\micron$ detection allows us to make a higher
resolution image using the visibilities of the longest baselines.
The resulting $3\sigma$ unresolved image provides 
a firm upper limit to the source size of 
$\lesssim 1\farcs2$ (10~kpc at $z = 1.325$), which is smaller than 
the size constraint given by the $\sim 1\farcs5$ VLA 1.4~GHz resolution.

The map of the test QSO 1419+543 
after adding the two sidebands and two tracks
is shown in Figure~\ref{fig1}~(d)~(inset).
A point source fit to the visibilities gave a 
positional offset from the phase center of 
$\Delta \alpha = 0\farcs11 \pm 0.05$ and $\Delta \delta = 0\farcs23 \pm 0.05$ 
for 1419+543. 
As with  1048+717 for GN~20, we fit a point source model 
to the visibilities of 1419+543 in each sideband of each track separately.
The results show that the offsets from the phase center 
had a wide range of values
($\Delta \alpha = 0\farcs04 -0\farcs28$ and $\Delta \delta = 0\farcs13 - 0\farcs25$)
with the overall average offset from 
the phase center of $\Delta \alpha = 0\farcs14 \pm 0.11$ and 
$\Delta \delta = 0\farcs21 \pm 0.05$.
These uncertainties are larger than those found in 1048+717 for 
GN~20, and are slightly larger than the uncertainties expected
from baseline errors.  It is possible that other factors such as image smearing
due to large phase noise may have introduced a small error in 
the source positions.
Thus, we assess an uncertainty 
of $\sim 0\farcs15$ in R.A. and $\sim 0\farcs2$ in Dec. for MIPS-J1428.

Figure~\ref{fig1}~(e) and (f) show the SMA $890\micron$ contours
of MIPS-J1428 overlaid on the NOAO Deep Wide Field Survey K-band and I-band
images.  The astrometry of both of these images is tied 
to the reference frame defined by the USNO-A2.0 catalog, 
and the typical rms of the residuals is $0\farcs35$ \citep{jannuzi05}.
The optical/NIR galaxy seen here is unambiguously
the galaxy aligned with the strong submm emission.  
The accurate astrometry of the submm emission provided by the SMA allows 
tight constraints on the separation between the bright optical/NIR position, 
and rules out fainter IRAC sources detected nearby \citep{borys05}.  Hence
MIPS-J1428 could lie directly behind the foreground $z=1.034$ galaxy, which 
would potentially result in a large amplification of the submm source.
\citet{borys05} use size/luminosity relationships to argue that despite
the alignment, the amplification is likely modest since the Einstein
ring is of comparable size to the known physical scales of local ULIRGs.  
Assuming that the lensing is modest, the size  
is comparable to that of other high-redshift submm sources \citep{chapman04}. 
Using the star-formation rate estimated
in \citet{borys05}, our limit on the angular size, $\theta$, of the object, 
we derive a lower limit on the SFR density of 
$>180(\theta/1.2)\mu^{-1}$M$_\odot$yr$^{-1}$kpc$^{-2}$ where $\mu$ is the
lensing amplification.  This is larger than
local ULIRGs \citep{meurer97} and comparable to other high-redshift 
submm galaxies \citep{chapman04}.

\section{Summary}

We present SMA observations of two 20~mJy submm sources, 
GN~20 and MIPS-J1428. The positions of the submm sources are 
determined with $0\farcs1 - 0\farcs2$ accuracy with these data, 
allowing precise identification of the correct
optical galaxy counterpart to the bright submm emission in GN~20, and
providing evidence that the foreground lens is only weakly affecting
the observed high FIR luminosity in MIPS-J1428.
If many of the  bright submm galaxies are slightly lensed objects 
similar to MIPS-J1428, then the implied source counts of 
inherently bright submm galaxies are over-predicted.
Detailed studies, however, exist for only a few sources, and
future surveys such as SHADES \citep{mortier05} 
will provide important information about 
the star formation properties at high-redshifts.

The Submillimeter 
Array is a joint project between the Smithsonian Astrophysical Observatory 
and the Academia Sinica Institute of Astronomy and Astrophysics, and is 
funded by the Smithsonian Institution and the Academia Sinica.
This work made use of observations made with the Spitzer 
Space Telescope, which is operated by JPL, 
California Institute of Technology under NASA contract 1407.
This work made use of images provided by the 
NOAO Deep Wide-Field Survey \citep{jannuzi99}
which is supported by the National Optical Astronomy Observatory (NOAO). 
NOAO is operated by AURA, Inc., under a cooperative agreement with the 
National Science Foundation.

\clearpage

\clearpage

\end{document}